\documentclass[]{aastex631}

\usepackage{amsmath} 
\usepackage{graphicx} 
\usepackage{epsf} 
\usepackage{float} 
\usepackage{csquotes}

\def\ucsc{Department of Physics, University of California, Santa Cruz, and Santa Cruz Institute for Particle Physics, Santa Cruz, CA 95064, USA}

\begin{document}
%\linenumbers
\title{Impacts of Voids,  Line of Sight Interactions, and Local Emission Environment on Detectability of Gamma-Ray AGN}

% Impacts of Voids, LOS interactions and local emission enviroment on detectability of gamma ray AGN

%Why So Voidy? Astrophysical Explanations for the Voidiness of Gamma-Ray Detected Galaxies

%Effects of Redshift, Local Emission Environment and Cascade Emission on Voidiness

% 

\author{Ollie Jackson}
\affiliation{\ucsc}

\author{Amy Furniss}
\affiliation{\ucsc}

\author{Olivier Hervet}
\affiliation{\ucsc}

\author{Megan Splettstoesser}
\affiliation{\ucsc}

\author{David A. Williams}
\affiliation{\ucsc}

\begin{abstract}
    Cosmic voids may have novel affects on the propagation of high-energy photons. We consider the fraction of the line of sight that intersect voids (termed \enquote{voidiness}). A previous study showed that active galactic nuclei (AGN) detected by \textit{Fermi} Large Area Telescope (LAT) lie along voidier lines of sight than redshift-matched populations of Sloan Digital Sky Survey (SDSS) optically detected quasars in the redshift range from $0.4 \leq z < 0.7$.  
    We explore this difference and various astrophysical explanations for it. Weaker intergalactic magnetic fields in voids would naturally enhance the gamma-ray cascading flux within the \textit{Fermi}-LAT point-spread function. We find that line-of-sight interactions increasing the flux in the \textit{Fermi}-LAT energy band by $\sim$0.1\% per Mpc of void traversed may be sufficient to result in the observed difference in voidiness distributions. Voidiness comparisons between SDSS QSO and AGN detected by imaging atmospheric Cherenkov telescopes at very-high-energies (VHE) do not yield any conclusive statement, likely because of the limited VHE sample size, and therefore are inconclusive about the role of possibly weaker extragalactic background light within voids. Finally, we measure that $28 \pm 3 \%$ of gamma-ray detected sources exist within a void (consistent with random mock populations) compared to $19.1 \pm 0.3 \%$ of SDSS quasars. We do not find any significant local void effect for gamma-ray sources that would explain the voidiness difference between \textit{Fermi}-LAT gamma-ray and SDSS QSO sources. These results suggest that the observed difference in voidiness distributions may be due to line-of-sight interactions rather than the local emission environment of gamma-ray AGN.
\end{abstract}

\section{Introduction}

Cosmic voids, the underdense regions between the cosmic filamentary structures of the Universe, track cosmic inhomogeneities, and thus allow for investigations of the homogeneity of cosmic fields, such as the extragalactic background light (EBL), as well as the intergalactic magnetic field (IGMF) and their effects on the propagation of very high-energy (VHE; E $\geq 100$ GeV) gamma-ray photons.  As gamma rays travel cosmological distances, they interact with EBL photons via pair production \citep{osti_4836265, 1967PhRv..155.1404G}, providing a means to probe the EBL density indirectly with observations of gamma rays from active galactic nuclei (AGN) (e.g. \cite{2015ApJ...812...60B, 2017A&A...606A..59H, 2018Sci...362.1031F, 2019ApJ...885..150A, 2019MNRAS.486.4233A, 2019A&A...627A.110B, 2024ApJ...975L..18G}). The electron-positron pairs produced in the photon-photon interaction provide opportunities to probe the IGMF via two main observational signatures. First, the pair-produced particles are deflected from their original photon path by the IGMF \citep{1994ApJ...423L...5A, 2009PhRvD..80l3012N, 2021Univ....7..223A}, which can result in the observation of a gamma-ray halo around a point source after inverse-Compton upscattering of photons from the cosmic microwave background (CMB) to gamma-ray energies \citep{halo1, halo2, halo3, halo4, halo5}. Second, even in the case that a halo is not sufficiently extended or bright enough for detection with gamma-ray instruments such as the \textit{Fermi} Large Area Telescope (LAT; \cite{halo4}) or ground-based imaging atmospheric Cherenkov telescopes (IACTs; e.g. VERITAS \citep{veritas}, HESS \citep{hess}, MAGIC \citep{magic}, and the future instrument CTAO \citep{ctao}), the pair-production and resulting cascade emission can reshape the observed gamma-ray emission spectrum (e.g. \cite{aharonian}).

The relative effect of these processes on the observed spectrum may be lessened if there is an overall lower density of EBL within cosmic voids, or if the magnitude of deflection of the charged pairs due to the IGMF within voids is insufficient to divert the pairs away from the original line of sight (LoS). Likely, the observed emission is changed from the intrinsic emission with a combination of both of these processes \citep{aharonian}. Additionally, specific environmental requirements for what makes a quasar emit gamma rays, beyond highly relativistic and beamed emission, are still a matter of relative uncertainty \citep{2020Galax...8...72C}. Understanding the role, if any, of the location of these gamma-ray emitters (both their local environmental conditions, e.g. inside or outside of an under-dense region of space, as well as the interactions which the gamma-ray photons undergo along their extragalactic path), is critical input for a complete understanding of these extreme objects. 

A recent study of the fraction of the LoS to the source that intersects voids (a quantity referred to as voidiness, defined in \cite{furniss_2015} and utilized in \cite{furniss_2025}, hereafter referred to as F25) found that gamma-ray emitting AGN are detected along lines of sight with higher fractions of intersecting void as compared to optically detected quasars. More specifically, the study found that \textit{Fermi}-LAT-detected gamma-ray sources lie along lines of sight with higher voidiness as compared to redshift-matched mock galaxy populations with random right ascension (RA) and declination (DEC) spatial distributions drawn from the population of optically detected quasars in the redshift range from $0.4 \leq z < 0.7$. This same result was not found with significance for sources in the local redshift range $0.1 \leq z < 0.4$ \citep{furniss_2025}. Initial studies by \cite{furniss_2015} indicated that EBL density may be marginally lower within voids, which was supported in a recent paper by \cite{hassan}.

Within this work, we further investigate the voidiness of the populations studied in F25, namely the \textit{Fermi}-LAT-detected AGN and optically detected quasars from the Sloan Digital Sky Survey (SDSS). This study utilizes the same 4th \textit{Fermi}-LAT Catalog of Active Galactic Nuclei, 3rd data release (hereafter referred to as 4LAC, \cite{4LAC2}), SDSS quasars (QSOs, \cite{SDSSDR16}), and voids from \cite{SutterDS7, SutterDS9}, along with the same voidiness values calculated for each source as found in F25. Applying the analysis done here and in F25 to the void catalog from the Dark Energy Survey Year 3 results \citep{DESY3_voids}, which covers an independent part of the sky, was considered as a potential independent check of the results. However, as the voids in \cite{DESY3_voids} are determined via a 2-D void-finding algorithm that uses photometric redshifts, and the Sutter void catalogs are produced with a 3-D algorithm and spectroscopic redshifts, a careful study beyond the scope of this work is required to consider the most appropriate way to compare the results extracted using the different catalogs. 

This study is outlined as follows. In Section \ref{sec: initial tests}, the redshift dependence of the results in F25 is further considered by examining the total intersecting distance traveled through cosmic voids (Section \ref{sec: full void distance}) as well as the effect of looking at different ranges of redshifts (Section \ref{sec: redshift ranges}). F25 noted three possible explanations for the observed difference in 4LAC and SDSS voidiness distributions, namely, (1) the possibility of weaker IGMF within voids, (2) decreased EBL density within voids, and (3) the local emission environment of gamma-ray sources.  We %then 
consider field interactions along the LoS (such as the deflection of pair-produced particles by the IGMF) in Section \ref{sec: sutter flux}, studying the possible interpretation that the voidiness, specifically the absolute distance that gamma-ray photons pass through voids, can change the intrinsic gamma-ray flux as reported in the 4LAC catalog \citep{4LAC2}. In Section \ref{sec: tev sources}, we consider the effect of large cosmic voids on the propagation of VHE-detected sources, and whether a decreased EBL density within voids may be the cause of the difference in voidiness distributions at distant redshifts. Finally, in Section \ref{sec: sutter in voids} the local emission environment of gamma-ray sources and whether gamma-ray emission is enhanced within voids is considered. We present our summary and conclusion in Section 6.  

\section{Intersecting Void Distance and Dependence on Redshift Range}\label{sec: initial tests}

\subsection{Investigation of Total Void Intersecting Distance} \label{sec: full void distance}

The analysis from F25 is repeated, replacing voidiness with the total intersecting distance through voids (in Mpc) for each source to assess if there is an equivalent difference in the 4LAC and SDSS QSO distributions of  intersecting void distance. With the voidiness determined as the ratio of the sum of line segments that intersect voids along the LoS to the total comoving distance to the galaxy, the total intersecting void distance is determined by multiplying the voidiness by the total comoving distance to the source.  
As in F25, we create 500 distributions of SDSS QSOs matched in redshift to 4LAC samples by drawing quasars at random from the SDSS sample to replicate the 4LAC redshift distribution.  We do statistical comparisons between the 4LAC samples and each of the redshift-matched SDSS samples, reporting the median results.  We note the possible bias introduced when studying distance-dependent aspects of BL Lacs, since a significant fraction of BL Lac objects do not have measured redshifts, as summarized in \cite{2015MNRAS.450.2404G}. 

\begin{figure}
    \centering
    \includegraphics[width=\linewidth]{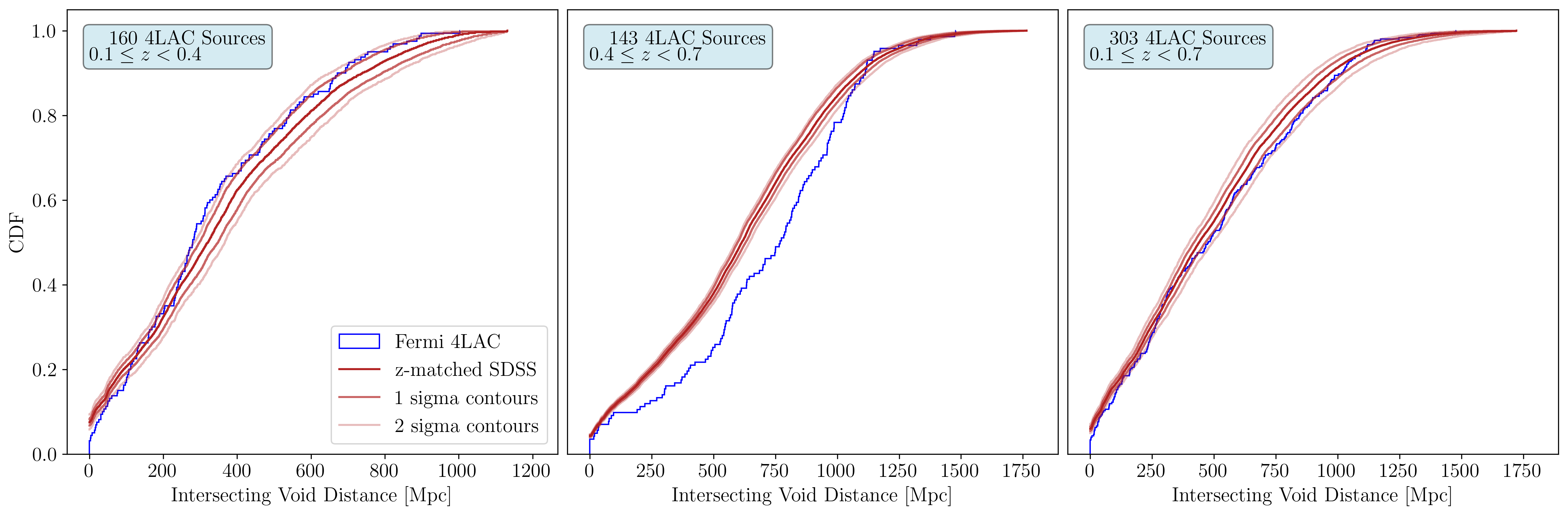}
    \caption{CDFs of the intersecting void distance distributions of 4LAC sources (blue) and the mean intersecting void distance of redshift-matched SDSS QSOs (in red) along with one and two standard deviations from the mean. The left panel is for sources in the redshift range $0.1 \leq z < 0.4$ (containing 160 sources), the center panel for $0.4 \leq z < 0.7$ (containing 143 sources), and the right panel for $0.1 \leq z <0.7$ (containing the total 303 sources).}
    \label{fig:full_void_mpc}
\end{figure}

When computed for the physical distance to sources, the median KS-statistics and median p-values (4LAC vs SDSS QSOs) are 0.084 and 0.327 for the redshift range of $0.1 \leq z < 0.4$ and 0.179 and $2.3 \times 10^{-5}$ for the redshift range of $0.4 \leq z < 0.7$, respectively, which is consistent with the previously reported results when comparing the population voidiness distributions. For completeness, the KS statistic and median p-value for the comparison of the distributions over the full redshift range $0.1 \leq z < 0.7$ are 0.077 and 0.097, respectively. These are again consistent with the comparison of population voidiness values over the full redshift range (see Section \ref{sec: redshift ranges}). Figure \ref{fig:full_void_mpc} shows the cumulative distribution functions (CDFs) for the intersecting void distances for the aforementioned cases.

\subsection{Voidiness of 4LAC and SDSS QSOs in Different Redshift Ranges}\label{sec: redshift ranges}

In F25, voidiness distributions for the 4LAC sources and SDSS QSOs were compared in two redshift ranges, with the intervals set to equally split the available redshift range of 0.1 to 0.7. A $4.1\sigma$ difference was found in the redshift range from $0.4 ~\leq z < ~0.7$. When comparing the populations in the full redshift range of $0.1 ~\leq z <~0.7$ with the same statistical tests, the two populations have similar voidiness distributions at the level of $1.6\sigma$.   %, corresponding to a significance of $1.62 \sigma$. 
%This indicates that the two voidiness distributions are not significantly different when we consider the total redshift range, and thus provides additional evidence, beyond the two-binned results provided in \cite{furniss_2025}, that any difference in voidiness between populations is specific to the redshift range tested. 

\begin{figure}[ht!]
    \centering
    \includegraphics[width = 0.5\linewidth]{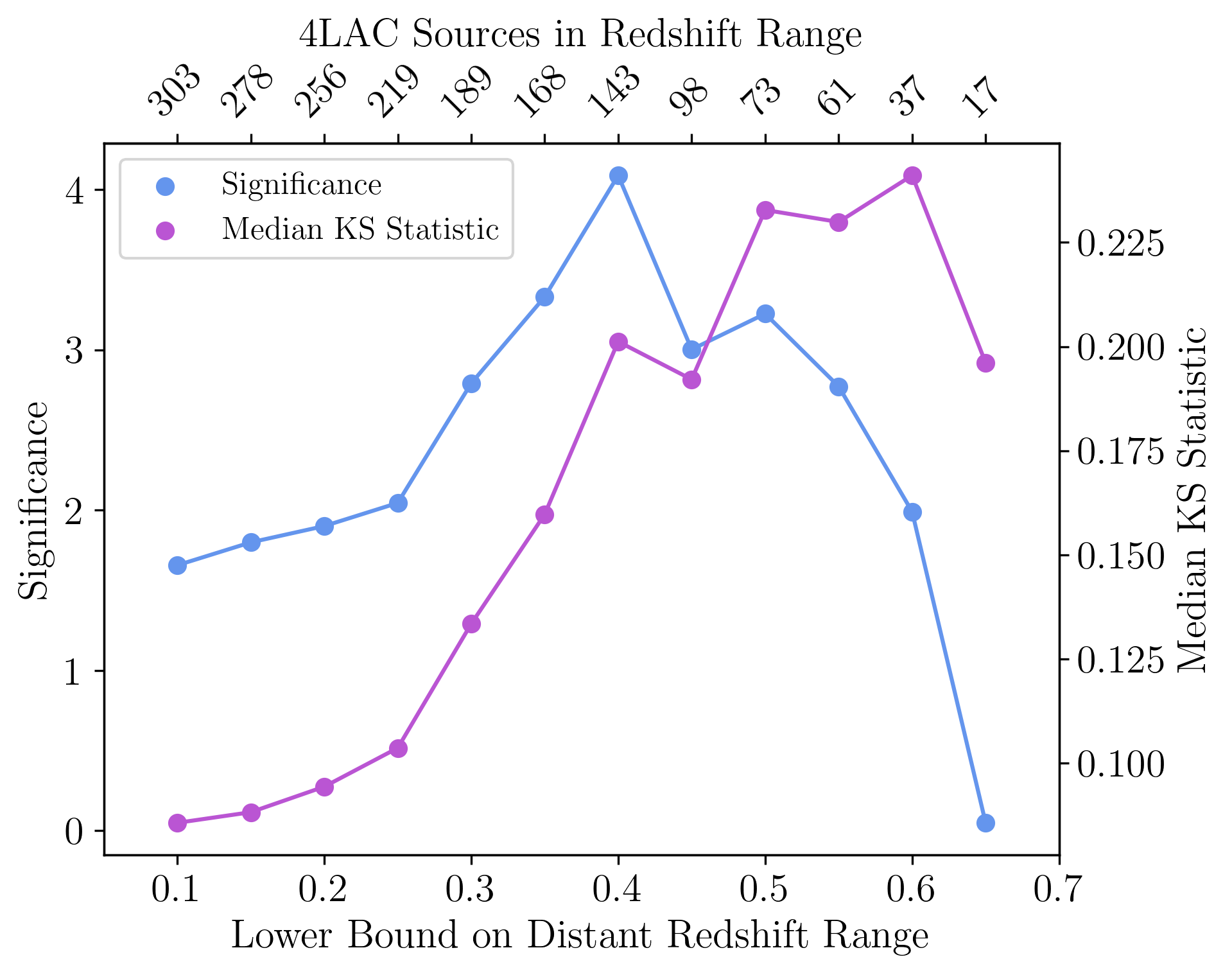}
    \includegraphics[width = 0.45\linewidth]{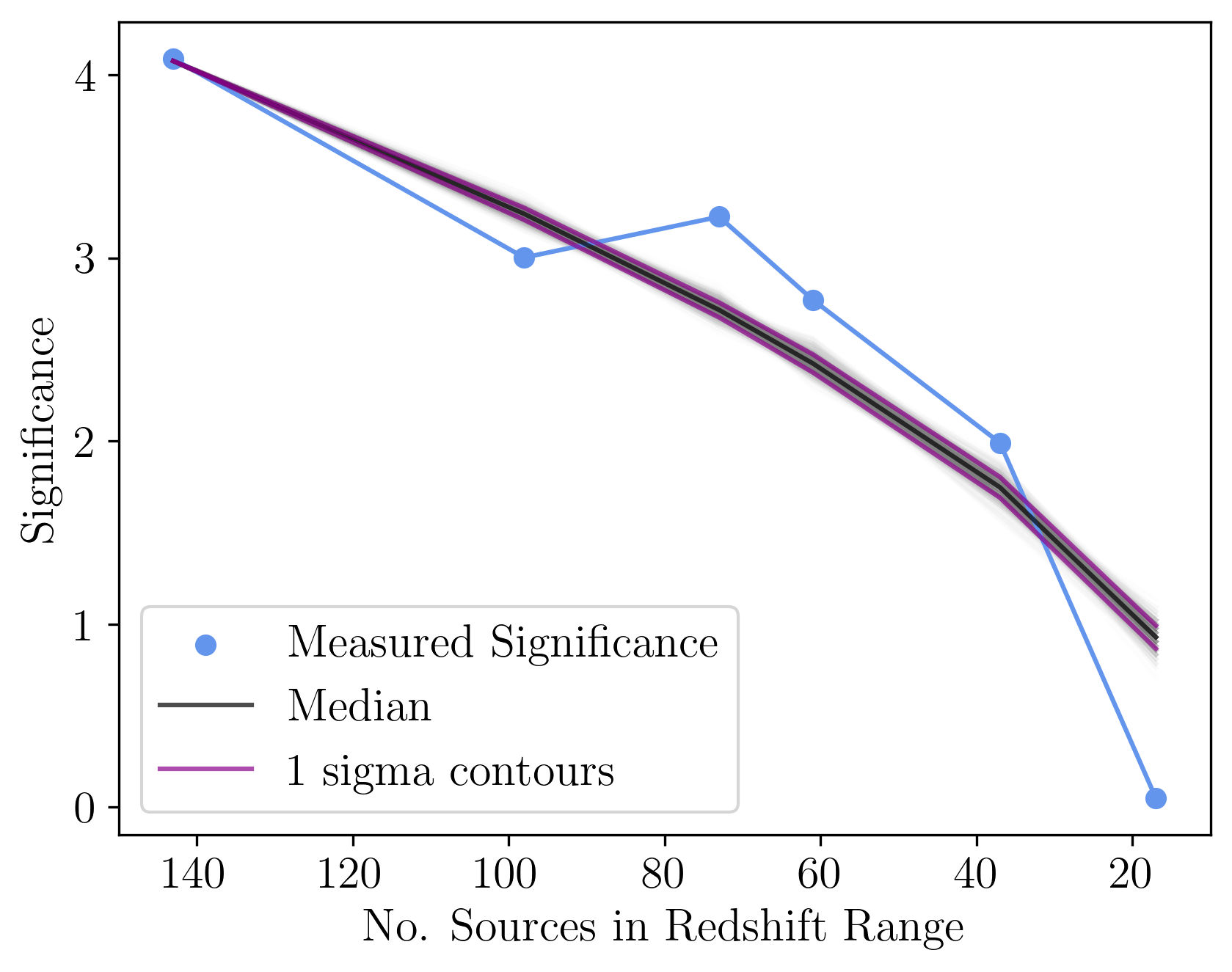}
    \includegraphics[width=0.5\linewidth]{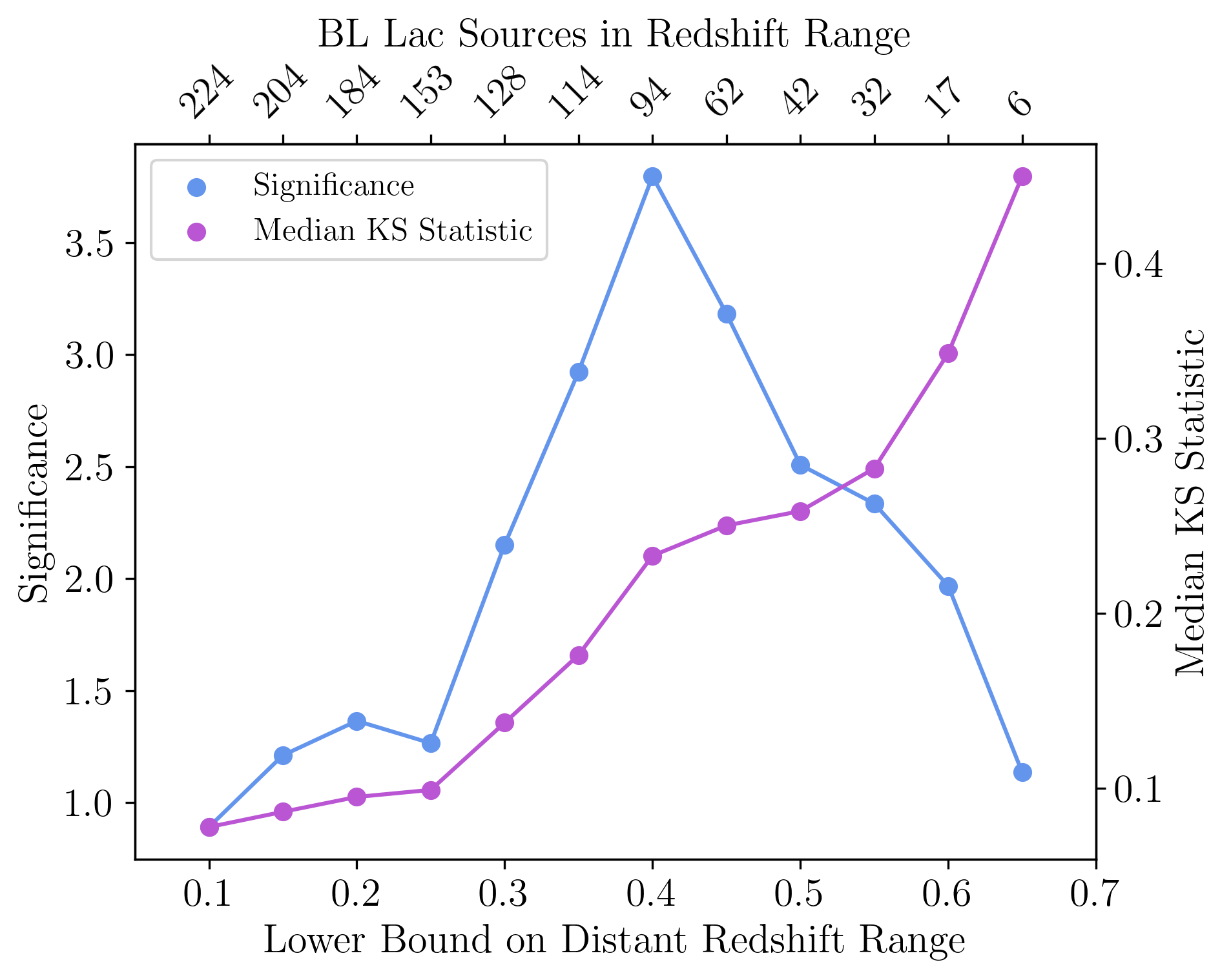}
    \includegraphics[width = 0.45\linewidth]{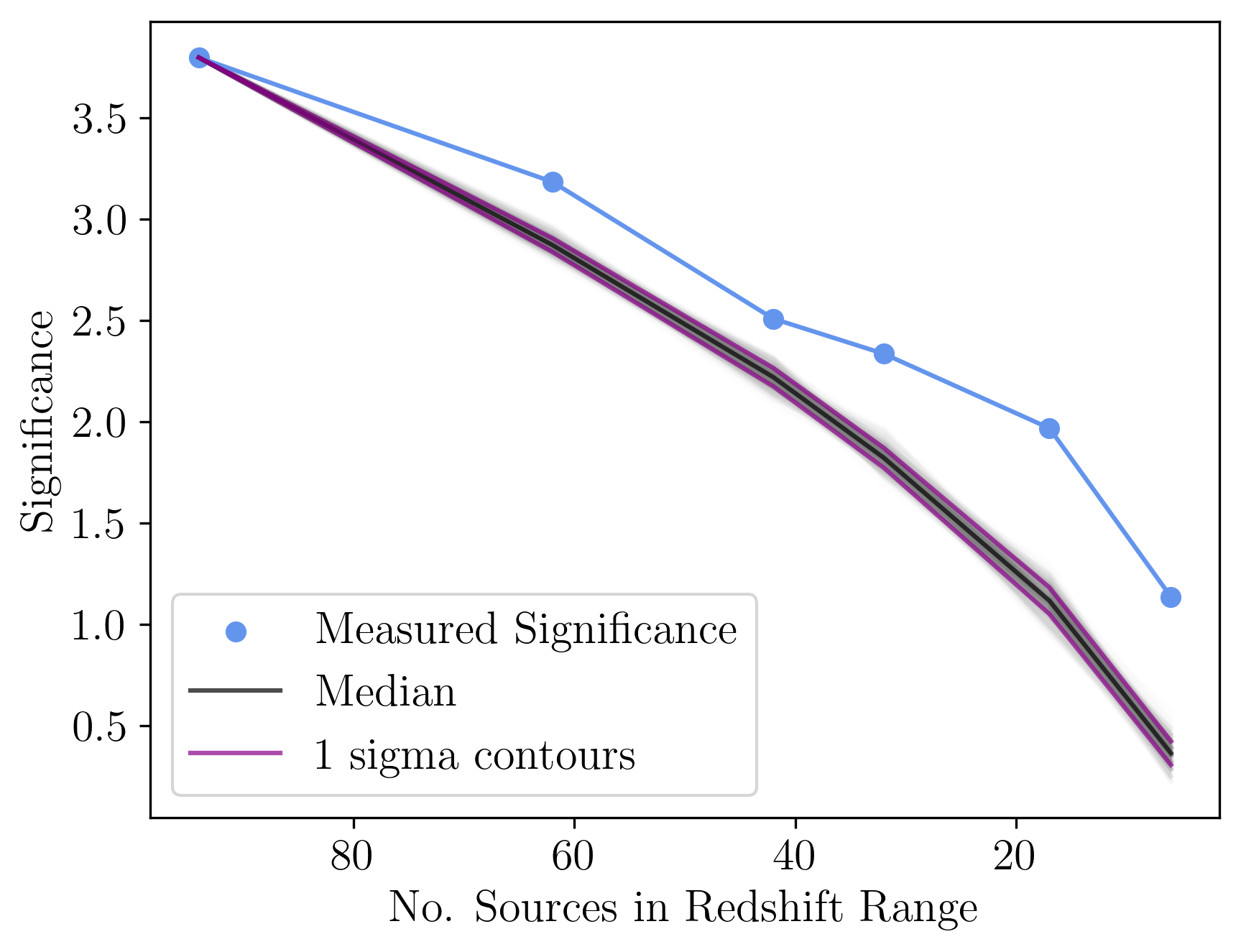}
    \caption{Top row displays results for 
    %only BL Lac sources, bottom for all 4LAC sources in sample. 
    all 4LAC sources in the sample; bottom row for only BL Lac sources. 
    Left: Significance of the difference between the voidiness distributions of 4LAC sources and SDSS QSOs (blue) and the KS statistics comparing the voidiness distributions of the two populations (purple). Right: For $z \geq 0.4$, the measured significance of the difference in voidiness distributions (blue) with the significance for 500  4LAC and SDSS QSO populations randomly removing sources (grey) to match the true population size in each redshift range. The median (black) and 1 sigma contours (purple) are also shown. }
    \label{fig:sutter_bin_change}
\end{figure}

We perform the comparison of voidiness distributions for different redshift ranges by setting the lower bound on the redshift interval which extends to $z = 0.7$. The left column in Figure \ref{fig:sutter_bin_change} displays the tension from the hypothesis that the two populations' voidiness distributions are derived from the same parent population, as measured with the KS test and reported in significance, as a function of the redshift range lower limit. The maximum significance of %$3.8\sigma$ (BL Lacs) and 
$4.1\sigma$ comes from from the redshift interval $0.4 ~ \leq z < ~ 0.7$, which, coincidentally, was the range used in F25. There is some variation in the statistical significance between the 4LAC and SDSS populations and in the KS statistic, but there is a definitive decrease in significance as the lower bound on the redshift range increases above $z = 0.4$. 

The KS statistic generally trends upward with increasing lower bound on the redshift range, in which the number of SDSS QSOs and 4LAC sources in the sample decreases and results are derived from increasingly limited statistics. For decreasing sample sizes, the KS test becomes less sensitive and is more likely to fail to detect differences in distributions \citep{KS_stat}. Thus, the fact that the KS statistic continues to increase even for decreasing sample sizes suggests that the difference between the voidiness distributions may persist, or even increase, as the lower bound is increased from $z = 0.4$, but that the lack of sufficient sources in these ranges inhibits sensitivity to the difference. 

In an attempt to assess whether the decrease in significance for the ranges with redshift lower bounds above 0.4 is due to the decrease in the number of sources, we simulate the effect of such a decrease and compare it with our results. We randomly remove sources from the 4LAC population until it has the same number of sources as the real population in each reported redshift range. For this test, it is only enforced that the remaining, randomly-chosen 4LAC sources have a lower redshift bound of $z = 0.4$. We start with the full 4LAC population and the redshift-matched population of SDSS QSOs, each from $0.4 \leq z < 0.7$. The KS test comparing the voidiness distributions of the random populations is repeated 500 times, and the significance distributions for all 500 populations are shown in the right column of Figure \ref{fig:sutter_bin_change}, along with the measured significance distribution. %The measured significance is above the median and $1\sigma$ ranges of the 500 random populations, from which we can interpret that the decrease in significance is due to the decreasing number of sources, and has no additional physical explanation. 

Additionally, F25 found that BL Lacertae (BL Lac) sources %overwhelmingly 
contribute to the majority of the difference in voidiness distributions: for the 94 BL Lacs, 27 flat-spectrum radio quasars (FSRQs), and 17 blazar candidates of unknown type (BCUs) the significance of the difference in voidiness distributions was $3.8\sigma, 0.3\sigma, \text{and} ~  2.7\sigma$, respectively. We use this as motivation to conduct the same investigation on the BL Lac sources independently of the complete 4LAC catalog. The corresponding plots are shown in the bottom panels of Figure \ref{fig:sutter_bin_change}. For the BL Lac population, the maximum significance of $3.8\sigma$ also occurs in the redshift range $0.4 \leq z < 0.7$ used in F25. The KS statistics, notably, continue to increase, reaching a maximum of 0.45 for $0.65 \leq ~ z < ~0.7$ for BL Lacs, which is measured however with only 6 4LAC BL Lacs, so the significance is less.

For the BL Lac sources, the measured significance of the real population as compared to the statistically reduced populations 
(shown in the lower right panel of Figure \ref{fig:sutter_bin_change})
is systematically higher, suggesting that the real voidiness discrepancy between 4LAC BL Lac and SDSS QSO sources might be higher than $3.8\sigma$ for redshifts $z > 0.4$ if more 4LAC BL Lac sources were detected. This effect is not seen when considering the full 4LAC sample, yet there is also no sign that the peak significance would decrease with higher redshift if we had higher statistics. In both cases, our results show that most of the observed significance decrease for $z > 0.4$ is due to the statistical decrease of 4LAC detected sources at higher redshifts.

%creating mock populations of 4LAC BL Lac sources with the same voidiness distribution in each redshift interval for $z \geq 0.4$, but maintaining the same number of mock sources as 4LAC BL Lac sources from $0.4 \leq z < 0.7$ (94 sources). Voidiness-matched mock 4LAC populations are created in an analogous way to the redshift-matched SDSS populations by constraining randomly populated voidiness values to fall within specific ranges based on the same bin edges of the true voidiness distributions.  We then conduct the same KS test comparing the mock 4LAC BL Lac populations to redshift-matched populations of SDSS QSOs for the same redshift ranges in Figure \ref{fig:sutter_bin_change}. Additionally, we place upper and lower error limits on the voidiness distributions based on the Poisson error of the number of mock sources in each voidiness bin, and additionally perform KS statistical tests comparing the upper and lower limit voidiness distributions to redshift-matched SDSS QSO populations in each redshift range. As shown in the right of Figure \ref{fig:sutter_bin_change}, we find that the significance distribution continues to increase within error in all higher redshift ranges, indicating that the observed difference in voidiness distributions of 4LAC BL Lac sources and SDSS QSOs may not actually peak at $z = 0.4$, but may continue to increase with more observed 4LAC sources at distant redshifts.

\section{Cascade Emission Contribution to Observed Gamma-Ray Flux}\label{sec: sutter flux}

\subsection{Void Characteristics and Connection to Intervening Fields}
As VHE photons travel extragalactic distances, they interact with the EBL via pair-production, resulting in charged pairs which can then be deflected by the IGMF before upscattering CMB photons to lower than VHE (e.g.\ GeV) energies.  
%Sources with lower EBL density would allow for more intrinsic VHE flux to traverse the extragalactic path.  
For a line of sight with weaker intervening IGMF, the upscattered GeV photons will have 
%a spatial extension spread tighter 
less lateral spread with respect
to the original VHE photon direction. 

The magnitudes of both the EBL and IGMF fields could be indirectly proportional, at least loosely, to void volume.  As the EBL is derived from the integrated starlight (emitted, absorbed, and re-emitted), it is reasonable to expect that the local density of the EBL would be lower in regions of space that lack galaxies (as explored in \cite{furniss_2015, 2017ApJ...835..237A, hassan}). Moreover, a natural connection exists between the magnitude of the IGMF and the large scale distribution of matter since charged matter seeds magnetic fields. This connection supports the possibility that the large underdense regions known as voids may also host, on average, lower magnitudes of magnetic field.  

Another way of stating the above connection is that baryonic density within a void is lower than average, and baryonic matter might trace %with 
the EBL and/or cosmological magnetic fields. The average size of detected voids increases with redshift, which is a known observational bias in the Sutter void catalog \citep{SutterDS7}. Additionally, \cite{SutterDS7} report the core density of voids, which we find decreases with increasing redshift for $z > 0.4$. 
%The average baryonic density within the voids reported in the Sutter void catalogs decreases with increasing void radius, as shown in \cite{SutterDS7}. Additionally, %as shown in Figure \ref{fig:sutter_void_size}%
%the average size of detected voids in the catalogs increases with increasing redshift, which is a known observational bias in the Sutter void catalog \citep{SutterDS7}. 
Combining these two distance-dependent void characteristics, the voids at higher redshift in our catalog have, on average, lower baryonic densities. If EBL density and IGMF strength do track with baryonic densities, more distant voids in the catalog would also have lower internal EBL density and IGMF strength.

\subsection{Void-related Flux Enhancement Estimations}

If photons propagate along a path with on-average weaker integrated IGMF, the cascade emission propagates more closely to the original VHE photon path. This cascade emission would contribute to an enhancement of the observed flux in the \textit{Fermi}-LAT sensitivity range as compared to the intrinsic emission in the same range. In order to probe the potential 
%magnitude 
effect of the cascade emission contribution, we artificially decrease the 4LAC measured flux in proportion to the Mpc distance of void through which the gamma rays travel, effectively removing 
possible
cascade emission from the observed flux. We refer to this lower flux as the cascade-corrected 4LAC flux. Following from the coordinates and spectral index of each 4LAC source, we use the sensitivity map associated with the 4LAC catalog (\texttt{detthresh\_P8R3\_12years\_PL22.fits}, \cite{sensitivity}) to determine the threshold flux at which sources would go undetected by \textit{Fermi}-LAT and therefore would not be reported in the 4LAC catalog. 4LAC sources with cascade-corrected fluxes that fall below the LAT sensitivity limit are removed from the sample. With the cascade-corrected 4LAC population, new redshift-matched SDSS populations are created in the same way as in F25, and the same statistical analysis is performed to determine if the cascade-corrected 4LAC population has a voidiness distribution that is significantly different than that of the SDSS redshift-matched QSO population. This is repeated for cascade-correction percentages of 0.005\% to 1\% per Mpc. Below a cascade correction of $0.005\%$ per Mpc, none of the fluxes of 4LAC sources drop below the LAT sensitivity, and for cascade corrections larger than $1\%$ per Mpc the sample 4LAC size becomes limited to less than 10. 

\begin{figure}[ht!]
    \centering
    \includegraphics[width = 0.5\linewidth]{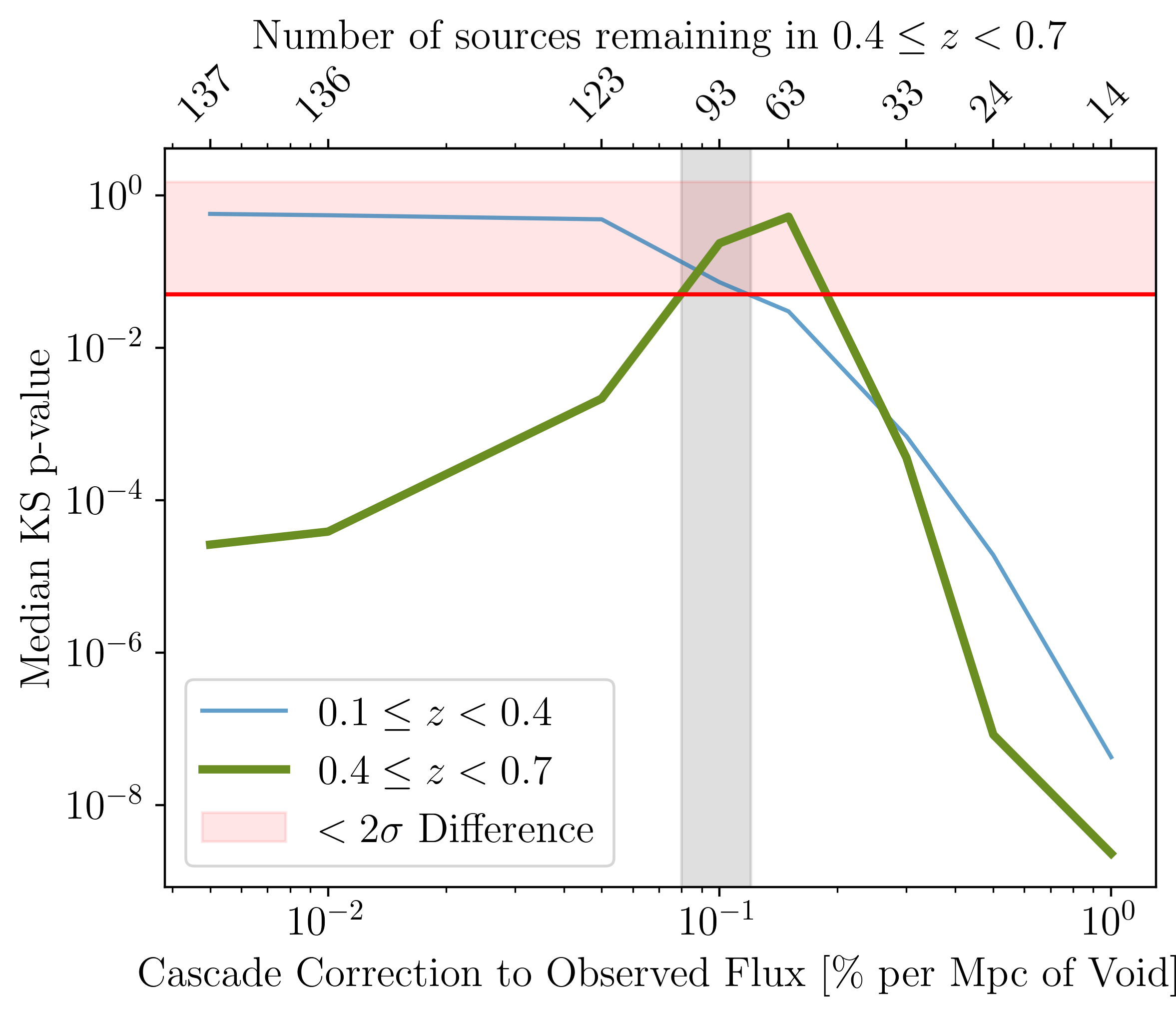}
    \caption{Median KS p-values resulting from the KS comparison of voidiness distributions of cascade-corrected 4LAC gamma-ray sources with redshift-matched SDSS QSOs as a function of the assumed flux correction percentage per Mpc of void intersection, shown for both the nearby redshift range from $0.1 \leq z < 0.4$ (in blue) and the distant redshift range from $0.4 \leq z < 0.7$ (in green). The number of sources which remain after the removal of sources which fall below the sensitivity range of LAT is noted on the top axis for $0.4 \leq z < 0.7$. The gray band corresponds to the region where the voidiness distributions in each redshift range for 4LAC and SDSS QSOs are consistent within $2\sigma$, which is for a cascade correction of approximately $(0.10 \pm 0.02)\%$ per Mpc of void intersection. Uncertainty is estimated based on the region in which the two populations have consistent voidiness distributions in each redshift range.}
    \label{fig:sutter_fluxpvals}
\end{figure}

The median p-value versus cascade-correction flux percentage is shown in Figure \ref{fig:sutter_fluxpvals}, demonstrating that for the two redshift ranges, the voidiness distributions of 4LAC and SDSS QSO sources are consistent with each other $(< 2\sigma ~\text{discrepancy})$ at a cascade correction of $(0.10 \pm 0.02)\%$ per Mpc of intersecting void. 

\begin{figure}
    \centering
    \includegraphics[width=\linewidth]{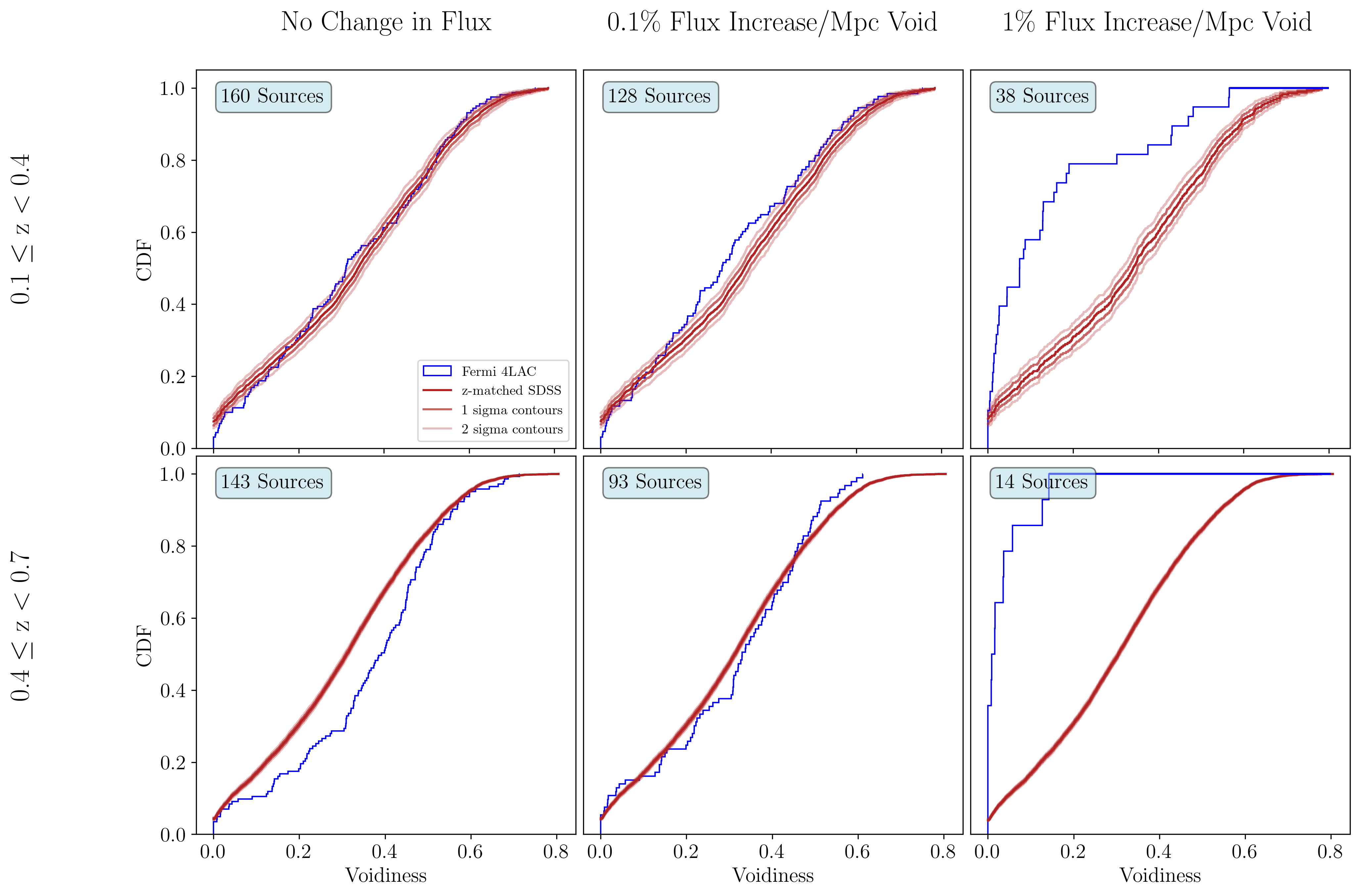}
    \caption{CDFs of the voidiness distributions in each redshift range for 4LAC sources (in blue) versus populations of redshift-matched SDSS QSO populations (in red) with contours for one and two standard deviations from the mean. From left to right the observed flux is cascade corrected by $0\%, ~0.1\%, \text{and} ~1\%$ per Mpc of void. The number of sources within each redshift range is provided in the upper left corner of each panel.
    We note that while cascade emission {\it increases} the observed flux compared to the emitted (intrinsic) GeV flux, to infer the detectability of AGN based on emitted flux alone, we {\it reduce} the observed flux.} 
    \label{fig:sutter_fluxtest}
\end{figure}

The correction of potential cascade emission as a percentage of Mpc intersection with voids is further visualized in Figure \ref{fig:sutter_fluxtest}, where, for both redshift ranges, the CDFs of three sample cascade-correction percentages are shown: the original reported fluxes from the 4LAC catalog, and cascade corrections of $0.1\%$  and $1\%$ per Mpc of intersecting void. In the nearby redshift interval, the 4LAC distribution (red) is consistent with the redshift-matched SDSS distribution before cascade corrections, and then at higher cascade corrections shifts the distribution to lower voidiness than the SDSS distribution. As 4LAC sources drop below the sensitivity of \textit{Fermi}-LAT, the sample becomes source-limited. 

For the distant redshift interval, the cascade-corrected 4LAC distribution starts by having a larger voidiness than the SDSS distribution, becomes consistent, and, as the cascade correction increases, changes to show a lower overall voidiness. The number of sources in each sample is reported in Figure \ref{fig:sutter_fluxtest}. As this cascade correction is calculated per Mpc of void that propagating photons intersect, sources in the distant redshift range will have more notable cascade corrections to flux. The sizable changes in voidiness distributions and significance seen in both Figures \ref{fig:sutter_fluxpvals} and \ref{fig:sutter_fluxtest} for $0.4 \leq z < 0.7$ are then expected. 

With the finding that 
LoS interactions at the level of $0.1\%$ per Mpc are sufficient to result in the observed difference in voidiness distributions from F25, 
%with a cascade correction to the observed flux being $0.1 \pm 0.002\%$, 
we investigate the median void path lengths in the two redshift ranges (280 Mpc and 770 Mpc), which correspond to median GeV flux corrections of $25\%$ and $54\%$ decreases, respectively. 
These corrections are neither so small as to be negligible nor so large as to be completely implausible.  
It is important to note that this cascade correction method serves only as a first order approximation of the behavior of high-energy photon propagation through voids.
Detailed calculations of the primary photon interactions and cascade emission production, including the primary photon spectrum produced by the AGN, will be needed to assess the viability of this scenario. We leave these calculations to a subsequent study.  
%This method inherently assumes that there is an infinite energy reservoir available to the photons to undergo cascades even after passing through many Mpc of void. For sources which emit VHE photons that propagate for large distances through voids, this correction likely does not occur across the entire extragalactic path length to the source, as there is not sufficient energy in intrinsic photons for cascade emission to continue indefinitely. Additionally, this study does not take into consideration the spectral index of the intrinsic gamma-ray emission from the source. Further studies which take into account the spectral index and total energy available for cascade emission are necessary, but are beyond the scope of this work. % as we focus on a simplistic model to determine if there is any region of the parameter space for which the voidiness distributions in the two redshift ranges are consistent.

%In addition to performing this flux analysis on the $0.1 ~ \leq z < ~0.4$ and $0.4 ~ \leq z < ~0.7$ redshift ranges, we also conduct the analysis for ranges with different lower bounds on the distant redshift interval (see right panel in Figure \ref{fig:sutter_fluxpvals}). As the lower bound on the redshift interval is increased, the voidiness distributions become more inconsistent, before turning around and becoming consistent again. This corroborates the result seen in the left of Figure \ref{fig:sutter_bin_change} of how the relationship between the two voidiness distributions changes with redshift bin. \textbf{Not sure what else to say about this plot}.

\section{Investigating EBL Density within Voids}\label{sec: tev sources}

Since photons with energy about 100 GeV can interact with EBL photons via pair production, the EBL density along the LoS can directly affect the extragalactic path length opacity. \cite{hassan} recently demonstrated that larger voids could harbor a lower local EBL density. They found that the optical depth at 13 TeV for the gamma-ray source GRB 221009A can be reduced by 10 percent and 30 percent for voids of radii 110 Mpc and 250 Mpc, respectively.  
%Due to the interaction of VHE photons with the lower-energy EBL photons, a lower EBL density within voids would translate to a lower gamma-ray opacity for sources with high voidiness. 
This lower VHE opacity resulting from lower EBL density in voids would translate to less absorption of gamma-ray photons. This result provides motivation for the investigation of the voidiness distribution of 4LAC AGN which have been detected at VHE energies by IACTs.  We refer to these IACT-detected AGN as VHE-detected for the remainder of this work.

As discussed in Section \ref{sec: sutter flux}, there is an observational bias reported in \cite{SutterDS7, SutterDS9} where the radius of the voids increases with redshift. Additionally, the density of voids decreases with increasing void radius. Therefore, in convolving these two trends, void density also decreases with increasing redshift of voids.  %We are motivated to consider the distance-specific characteristics of voids because if voids have an on-average lower EBL density, then the void distance through which photons travel would greatly impact the observed spectrum. 
The observation here and in F25 that the difference in voidiness distributions between gamma-ray detected and SDSS QSO sources is only significantly present in more distant redshift ranges encourages consideration of how voids and void density would impact the propagation of VHE photons at distant redshifts. 

Since gamma-ray opacity due to the EBL starts at approximately 100 GeV, we study the gamma-ray spectra and voidiness of the 16 VHE-detected sources within the 4LAC catalog that coincide with the void catalog footprint.  Of these, 14 are BL Lac sources and 2 are FSRQs. The spectrum reported in 4LAC, including the power-law (PL) index, variability index, source luminosity calculated from the 4LAC integrated photon flux between 1 and 100 GeV, and the voidiness for each of these objects is summarized in Table \ref{tab:tev_sources}. The power-law index represents the slope of the differential photon flux in the 1 to 100 GeV band. The variability index is an indicator of source variability, and is defined as twice the sum of the log-likelihood difference between the flux fitted in each time interval and the average flux over the full catalog integral \citep{4LAC}. Sources are noted as non-variable if they have a variability index of less than 24.725, which corresponds to a $1\%$ chance of the source showing significant variability \citep{variability}. 

In addition to the VHE sources noted within the 4LAC catalog with the \texttt{TeVflag} = 1, all recently announced VHE-sources in TeVCat\footnote{https://tevcat.org/} are checked to ensure that all sources detected at VHE energies since the publication of 4LAC are also included in this study. No additional VHE sources were added to our sample. %This check resulted in the identification of one additional VHE-detected source within the void footprint, namely 4FGL J1136.8+2550 (RX J1136.8 + 2551). 

\begin{table}[]
    \centering
    \begin{tabular}{|c|c|c|c|c|c|c|c|}
    \hline
    Source Name & TeVCat Name & Class & Voidiness & Redshift& PL Index & Luminosity  & Variability \\
    & & & & & &$(\times 10^{46})$ [ph/s]& Index\\
    \hline
    4FGL J1428.5+4240 & H 1426+428 &  BL Lac & 0.006 & 0.129 & $1.65 \pm 0.04$  & $2.67 \pm 0.17$ &18.56\\
    4FGL J0847.2+1134& RBS 0723 &BL Lac & 0.074 & 0.199 & $1.73 \pm 0.06$ & $4.56 \pm 0.46$ &20.81\\
    4FGL J0854.8+2006& OJ 287 &BL Lac & 0.191 & 0.306 & $2.21 \pm 0.01$ & $165.0 \pm 3.4$ & 920.42\\
    4FGL J1422.5+3223 & B2 1420+32 & FSRQ & 0.231 & 0.682 & $2.21 \pm 0.01$ & $1034 \pm 20$ &9817.71\\
    4FGL J1015.0+4926 & 1ES 1011+496 & BL Lac & 0.303 & 0.212 &  $1.84 \pm 0.01$ & $109.6 \pm 1.7$ & 338.85\\
    4FGL J1058.6+2817 & GB6 J1058+2817 & BL Lac & 0.330 & 0.479 & $2.24 \pm 0.06$ & $42.2 \pm 3.4$ & 48.25\\
    4FGL J1224.9+2122 & 4C +21.35 & FSRQ & 0.350 & 0.434  &  $2.34 \pm 0.01$ & $992 \pm 12$ & 25039.39\\
    4FGL J1442.7+1200 & 1ES 1440+122 & BL Lac & 0.364 & 0.163 & $1.75 \pm 0.05$ & $3.88 \pm 0.34$ & 21.07\\
    4FGL J1031.3+5053 & 1ES 1028+511 & BL Lac & 0.369 & 0.361 & $1.74 \pm 0.03$ & $45.0 \pm 2.2$ & 15.37\\
    4FGL J1555.7+1111 & PG 1553+113 & BL Lac & 0.375 & 0.433 &  $1.68 \pm 0.01$ & $1038 \pm 17$ &168.60\\
    4FGL J1224.4+2436 & MS 1221.8+2452 & BL Lac & 0.394 & 0.219 & $1.93 \pm 0.04$ & $16.67 \pm 0.85$ & 228.08\\
    4FGL J1221.5+2814 & W Comae & BL Lac & 0.471 & 0.102 &  $2.15 \pm 0.02$ & $8.68 \pm 0.23$ & 339.33\\
    4FGL J1417.9+2543 & RBS 1366 & BL Lac & 0.539 & 0.237  & $1.50 \pm 0.07$ & $3.69 \pm 0.58$ & 17.36\\
    4FGL J1217.9+3007 & 1ES 1215+303 & BL Lac & 0.565 & 0.130 &  $1.93 \pm 0.01$ & $44.69 \pm 0.70$ & 645.82\\
    4FGL J1221.3+3010 & 1ES 1218+304 & BL Lac & 0.567 & 0.184 & $1.72 \pm 0.02$ & $36.3 \pm 1.0$ & 116.60\\
    %4FGL J1136.8+2550 & BL Lac & 0.595 & 0.156 & $1.96 \pm 0.10$ & $1.50 \pm 0.21$ & 16.44 \\
    4FGL J1427.0+2348 & PKS 1424+240 & BL Lac & 0.611 & 0.604 &  $1.82 \pm 0.01$ & $1734 \pm 25$&370.87\\
    \hline
    \end{tabular}
    \caption{VHE-detected sources within the 4LAC catalog which are within the void catalog footprint, organized by increasing voidiness.  }
    \label{tab:tev_sources}
\end{table}

\begin{figure}
    \centering
    \includegraphics[width=0.6\linewidth]{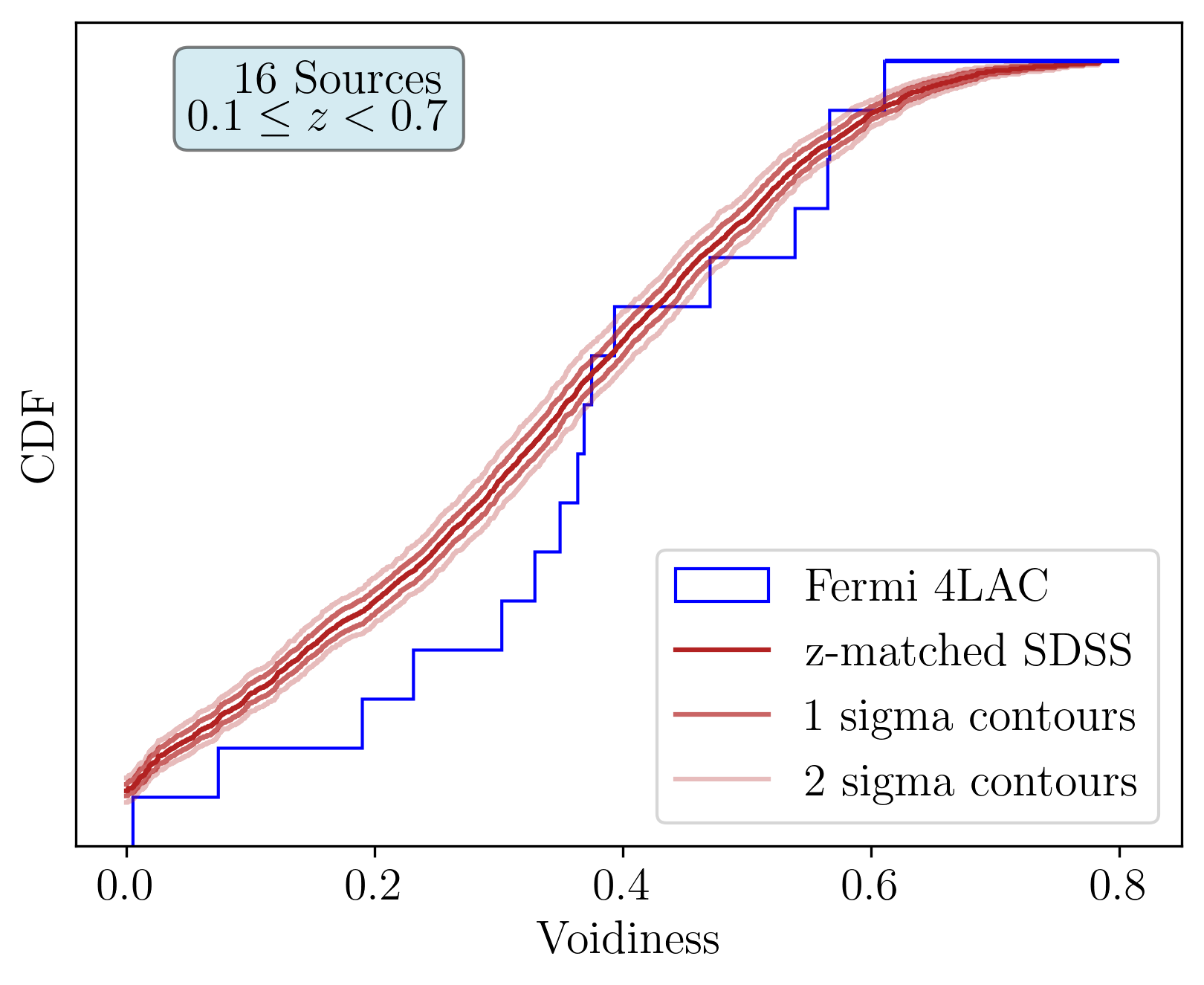}
    \caption{CDF of the voidiness distributions of 4LAC VHE-detected sources (in blue) and median voidiness of redshift-matched SDSS QSOs (in red), with the one and two standard deviation contours. The corresponding KS statistic and p-value are 0.23 and 0.31, respectively.}
    \label{fig:tev_cdfs}
\end{figure}

For VHE-detected blazars, we conduct the same analysis detailed in F25 to compare the voidiness distribution of the 4LAC VHE-detected sources to the voidiness distribution of redshift-matched populations of SDSS QSOs. %In the redshift range from $0.1 \leq z < 0.4$, we find a KS-statistic of 0.23 and median p-value of 0.48, while in the distant redshift bin from $0.4 \leq z < 0.7$ we find a KS-statistic of 0.34 and a median p-value of 0.50, showing that the voidiness of VHE-sources do not deviate in any significant way from the redshift-matched SDSS QSO populations.
The population has too 
%limited statistics 
few objects
to break it down into redshift ranges. Namely, there are only 11 4LAC sources in the redshift range $0.1 \leq z < 0.4$ and 5 in the redshift range $0.4 \leq z < 0.7$; therefore the analysis is only conducted in the full redshift range from $0.1 \leq z < 0.7$.  The KS statistic is 0.21 and the median p-value is 0.42. Figure \ref{fig:tev_cdfs} displays the CDF of the voidiness distributions. The VHE-detected population has a larger voidiness as compared to the redshift-matched populations, but with no statistically significant difference in overall voidiness distribution.

In an attempt to investigate the voidiness distribution for a larger population of AGN detected at VHE, we also compared the voidiness distributions of 
%the redshift-matched SDSS quasars with 
galaxies from the 3rd Fermi Hard Source List \citep{2017ApJS..232...18A} %. This 
with populations of SDSS quasars matched in redshift.  The Hard Source List
catalog contains sources detected above 10 GeV in the first 7 years of LAT data.  With 61 and 27 sources in the 0.1 $\le z < 0.4$ and $0.4\leq  z <0.7$ ranges, respectively, we see the sources in the nearby range to show no statistical difference with a p-value of 0.85 and KS-statistic of 0.08, while in the more distant range we see a marginal indication of different voidiness distribution with a p-value of 0.005 and KS-statistic of 0.32.

Motivated by the idea that, to at least some level, the \textit{Fermi}-LAT observed spectrum may have both intrinsic and cascade contributions that are dependent on the EBL and IGMF (as summarized in Section \ref{sec: sutter flux}), we look for correlations between the voidiness and gamma-ray spectral and variability characteristics of 4LAC sources. 
%For example, if voids do host lower overall IGMF magnitude, sources with lower voidiness (on average higher intervening IGMF) would display a lower level of variability since cascade particles and emission will travel along longer (angularly deflected) paths than intrinsic emission, washing out prompt variability. 
Such correlations could arise if the voidiness affects the relative contribution of cascade emission.
%Moreover, 
Cascade emission could contribute to the flux in the GeV band, altering the observed spectrum from the intrinsic emission in a way which depends on the LoS field density of EBL and IGMF. 
In addition, the variety of path lengths traversed by cascade emission spreads out its arrival, smoothing out variability compared to the intrinsic emission.
% spectral characteristics would vary (higher luminosity with lower IGMF/voidiness) based on the reprocessed flux contribution in the GeV band. Since the cascade emission affects the spectrum in the GeV band, it may impact the measured spectrum in a way that is correlated to line of sight density. 

We look for correlations within four specific populations: all 4LAC sources in the study, non-variable 4LAC sources (selected by 
%making a cut on variability 
removing 4LAC sources with a variability index above 24.725), all VHE-detected sources, and 4LAC sources that are within voids (see Section \ref{sec: sutter in voids}). Cutting on variability for VHE-detected sources leaves only six non-variable objects, which is too low to determine possible correlations and is therefore omitted. The Pearson correlation coefficient (\texttt{scipy.stats.pearsonr} Python function) is calculated for each of the aforementioned 4LAC-reported measurements of source characteristics with respect to the voidiness values taken directly from F25.  Each Pearson coefficient, reported in Table \ref{tab:tev_voidiness_comp}, is consistent with random Monte Carlo sampling, meaning that no significant correlations are found.   In addition to the Pearson coefficients, Spearman and Kendall coefficients were tested (\texttt{scipy.spearmanr} and \texttt{scipy.kentalltau}, respectively), both of which indicated no significant non-linear correlations.

\begin{table}[]
    \centering
    \begin{tabular}{|c|c|c|c|c|}
    \hline
     Spectral Parameter &All 4LAC Sources  & Non-Variable 4LAC Sources & 4LAC VHE-detected Sources & 4LAC in voids  \\
    \hline
    No. Sources & 303 & 216 & 16 & 84\\
     Luminosity & 0.049 & -0.035 & 0.210 & 0.141\\
     Power-Law Index & -0.035 & -0.003& -0.100 &-0.055\\
     Variability Index & 0.005 & 0.082 & -0.080 & 0.042\\
     \hline
    \end{tabular}
    \caption{Pearson coefficients of correlations between voidiness and the source luminosity, power-law index, and variability index for four populations. (See Section \ref{sec: sutter in voids} for information about 4LAC sources within voids.) None of the Pearson coefficients indicates a significant correlation.}
    \label{tab:tev_voidiness_comp}
\end{table}

\section{Local Emission Environment of Sources}\label{sec: sutter in voids}

In F25, one of the explanations for why the voidiness distribution may be different between gamma-ray-detected and optically-detected sources is that voids may provide an environment that is advantageous for gamma-ray production. We investigate the fraction of 4LAC AGN that lie within voids and assess whether the difference in voidiness distributions between 4LAC sources and SDSS QSOs may be due to source location being inside or outside of a void.

A source was determined to be within a void if it lies within a sphere centered at the RA, DEC, and redshift of the void %(converted to 3d coordinates), 
with a radius given by the effective void radius, as reported in the void catalog \citep{SutterDS7, SutterDS9}. Of the 303 4LAC sources used in this study and in F25, 84 of them lie within a void. Broken down by the class of gamma-ray source, there are 9 FSRQs ($21\%$ of total FSRQs), 67 BL Lacs ($30\%$ of total BL Lacs), 1 narrow-line Seyfert 1 (NLSY1; $25\%$ of total NLSY1s), and 7 BCUs ($24\%$ of total BCUs). Likewise, of the 19751 SDSS QSOs, 3768 lie within a void. Thus, $28 \pm 3\%$ of 4LAC sources lie within a void, while $19.1 \pm 0.3\%$ of SDSS QSOs lie within a void. The error is taken as the square root of the number of sources in voids, estimating the population as a Poisson distribution. 

\begin{figure}
    \centering
    \includegraphics[width=0.45\linewidth]{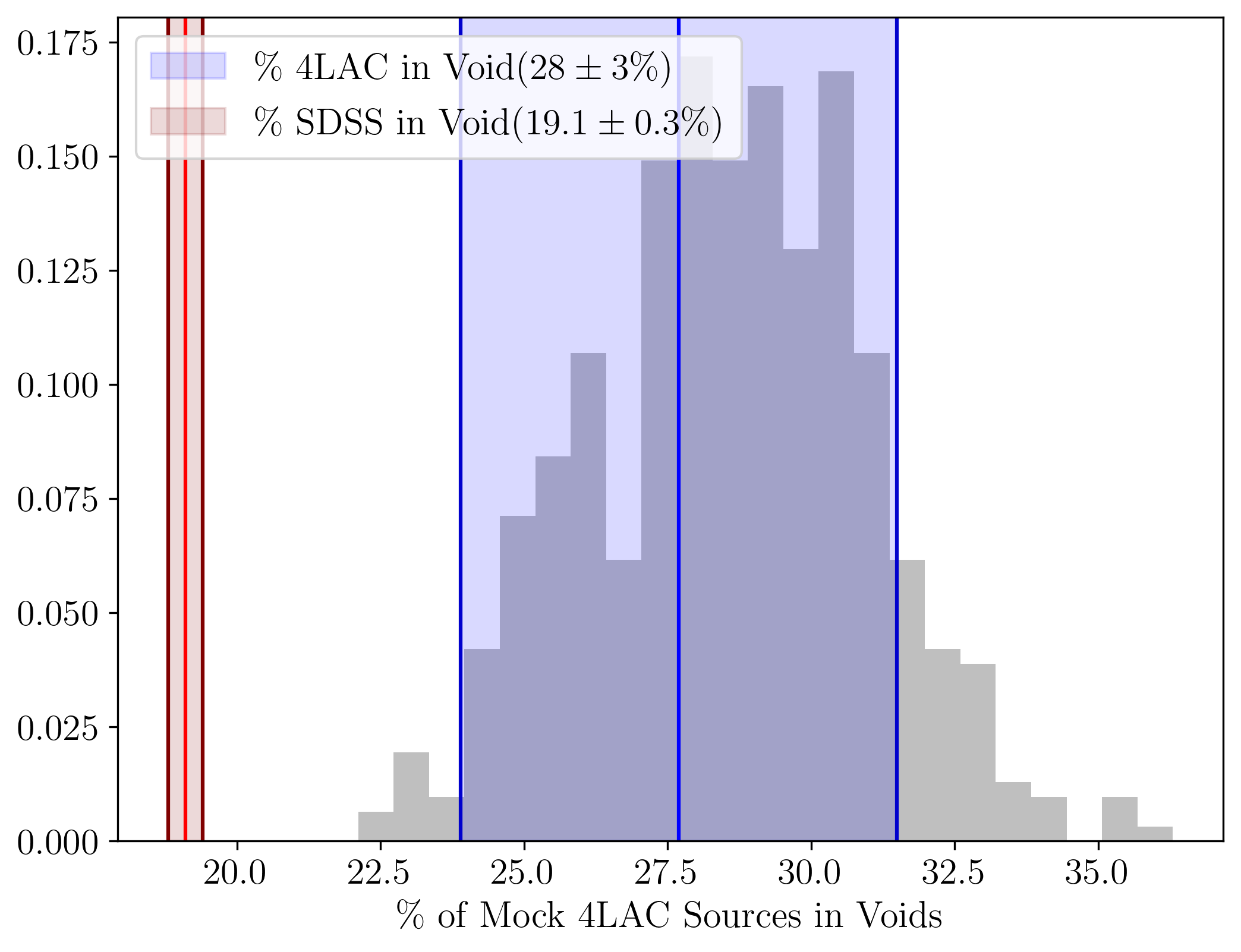}
    \includegraphics[width = 0.45\linewidth]{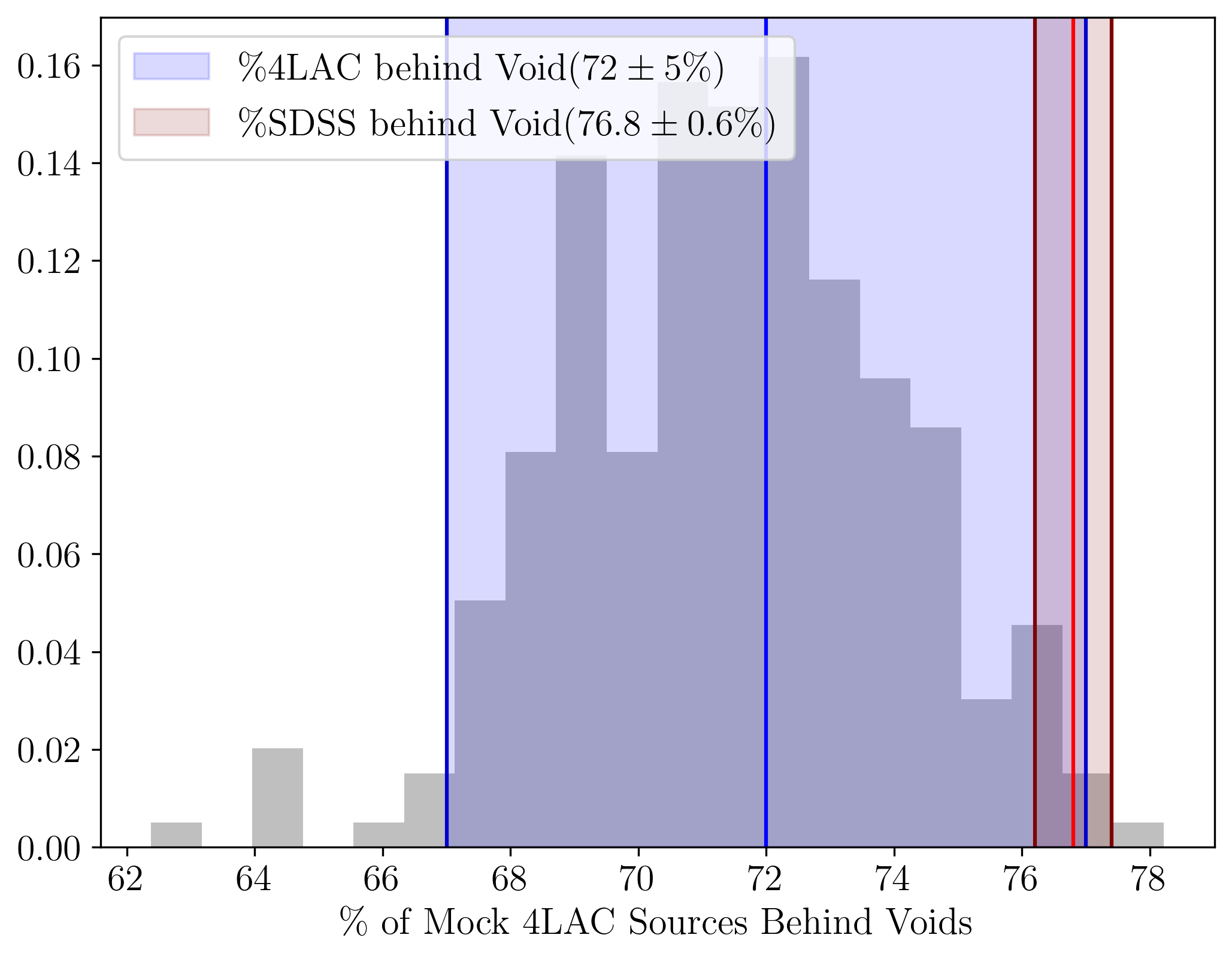}
    \caption{Left: Percent of redshift-matched mock 4LAC sources in voids. Right: Percent of redshift-matched 4LAC sources behind voids. Both panels display 500 simulated populations (normalized histogram, in gray) with the measured percentage of 4LAC sources with uncertainty in blue and measured percentage of SDSS QSOs with uncertainty in red.}
    \label{fig:sources_in_voids}
\end{figure}

We create 500 4LAC populations with the same redshift distribution as the cataloged 4LAC sources, but with randomly sampled RA and DEC locations within the Sutter void footprint. These 500 populations are hereafter referred to as mock 4LAC populations. Each mock population contains 303 sources, matching the total number of 4LAC sources. Using the Sutter void catalog, the percentage of mock 4LAC sources within voids is consistent with the measured percentage of 4LAC sources in voids. The percentage of mock 4LAC sources within voids is never within the measured percentage of SDSS QSOs in voids. These results are demonstrated in the left panel of Figure \ref{fig:sources_in_voids}. The measured percentage of 4LAC sources within voids is consistent with randomly distributed gamma-ray sources, while the measured percentage of SDSS quasars is not consistent with random distribution. This aligns with the result from F25 that 4LAC source voidiness was determined to be consistent with voidiness distributions of randomly located galaxies within the void footprint. 

We additionally consider the percentage of 4LAC and SDSS QSO sources that are behind voids (i.e. sources that are not in voids, but have non-zero voidiness, which equates to 209 out of 303 4LAC sources). 500 mock 4LAC populations are again created with the same redshift distribution as the true 4LAC population, and the percentage of mock 4LAC sources that are behind voids is determined. The results are shown in the right panel of Figure \ref{fig:sources_in_voids}. The percentages of true 4LAC and SDSS QSO sources behind voids are consistent with each other within Poisson error and are also both consistent with randomly distributed mock sources within the void footprint.

To determine if the difference between the number of 4LAC and SDSS QSO sources in voids significantly contributes to the observed difference in voidiness distributions, we look at the voidiness distribution of 4LAC sources as compared to SDSS QSOs after removing the contribution to the voidiness from the void that a source resides within. For sources in voids, the distance that the gamma rays propagate from the source until they exit a void is removed from the LoS voidiness fraction. %The KS test is then performed again comparing the \enquote{cut} voidiness distribution of the 30 4LAC sources in voids to the voidiness distribution of redshift-matched SDSS QSOs, resulting in the significance decreasing to $2.4\sigma$. 
The KS test comparing the voidiness of the total population of 4LAC sources to redshift-matched SDSS QSOs, replacing the voidiness values of both the 4LAC and SDSS sources in voids with the \enquote{cut} voidiness values which have been decreased by the void intersection distance of the void the source resides in, results in a p-value corresponding to a significance of $3.5\sigma$. Although this is a decrease in significance from $4.1\sigma$, a difference in voidiness distributions remains after this adjustment. Despite the observation that 4LAC sources are more likely to be within voids relative to SDSS QSOs, source location is not a primary reason for the difference in the SDSS QSO and 4LAC voidiness distributions.

\section{Conclusion}

We extend the study of voidiness distributions determined in F25.
%investigating the possibility that the difference found for \textit{Fermi}-LAT-detected AGN and SDSS detected quasars may be derived from weaker IGMF within voids, decreased EBL density within voids, and/or the local emission environment of gamma-ray sources.
Using the intersecting void distance to the objects rather than the voidiness, which is the fraction of void distance, yields a difference between the 4LAC and SDSS samples with essentially the same statistical significance as F25.
The difference between the voidiness distributions of 4LAC and SDSS QSO sources is most significant for the redshift interval $0.4 \leq z < 0.7$ (as used by coincidence in F25), compared to other intervals with lower bounds between 0.1 and 0.65. The reduced significance in redshift ranges with lower bounds above 0.4 is consistent with arising entirely from the reduced number of objects in the sample, and does not imply a reduction in the effect at larger redshifts.
%By determining the significance in difference of voidiness distributions with randomly removed sources, we interpret that the decrease in significance at larger redshifts is likely mostly due to a decrease in sample size. 
These investigations support the conclusion of F25, but do not tease out a more significant statistical difference between the 4LAC and SDSS samples.
%Furthermore, the difference between voidiness distributions and the difference between intersecting void distance distributions for 4LAC and SDSS QSO sources are consistent, which supports the hypothesis that the difference may arise from line-of-sight interactions that would depend on an integrated path density. Here, voidiness and intersecting void distance act as proxies for the integrated path density. 

When the 4LAC flux is artificially corrected for a potential contribution from cascade emission by decreasing the observed flux by 0.1$\%$ per Mpc of void along the LoS, the voidiness distributions of cascade-corrected 4LAC sources and SDSS QSOs differ by a significance of less than $2\sigma$. Hence, it is possible that SDSS QSOs and gamma-ray emitting AGN could have similar underlying voidiness distributions (and similar spatial distributions in general), but that gamma-ray emitting AGN which exist along lines of sight with higher void intersection are more likely to be detected as they have a larger portion of cascade emission reaching Earth. Additional simulation work, beyond the scope of this current study, will be needed to establish whether this increased contribution from cascade emission in the GeV band is plausible.  

Considering that the EBL density within voids is, on average, lower,
a large voidiness could favor VHE detection.
%and by 
Looking at VHE-detected sources specifically, there is no significant difference in voidiness distributions between 4LAC VHE sources and SDSS QSOs. We note, however, that given the small VHE-detected sample of only 16 sources, there is limited sensitivity to measuring a difference between the two populations.

Finally, when considering the location of gamma-ray sources as being within or behind voids, the percentage of 4LAC sources within voids is consistent with randomly distributed sources, and is larger than the percentage of SDSS QSO sources in voids. However, this difference in the percentage of sources in voids does not appear to be responsible for the observed difference in voidiness distributions,
since a difference in the voidiness distributions of the 4LAC and SDSS sources remains even after removing the contribution from the void (if any) in which each object resides.

The studies presented here confirm the findings of \cite{furniss_2025}, but are not able to achieve higher statistical significance.  They are suggestive that the origin of the voidiness difference between AGN detected by the \textit{Fermi}-LAT and quasars detected by the Sloan Digital Sky Survey is related to line-of-sight interactions as gamma-rays propagate to Earth, rather than a preference for gamma-ray production in or near voids.
However, further investigation of the details of the gamma-ray propagation effects is needed to learn whether they can accurately account for the difference.

\section{Acknowledgments}

We thank the referee for their insights and suggestions. We would like to thank Andras Kovacs for valuable discussions about this work. We additionally thank Josepf Amador for his work in \cite{furniss_2025} that influenced this study. We gratefully acknowledge support from the U.S. National Science Foundation awards PHY-2110974, PHY-2310002, and PHY-2411760.

\bibliography{biblio}{}

\bibliographystyle{aasjournal}

\end{document}